 \documentclass[aps,twocolumn,showpacs,nofootinbib,nobibnotes,superscriptaddress]{revtex4}

\usepackage{epsfig}
\usepackage{amssymb}

\begin{document}


\title{On the nature of incompressible magnetohydrodynamic turbulence}

\date{\today}

\author{G. Gogoberidze}
\affiliation{Georgian National Astrophysical Observatory, 2a Kazbegi
ave., 0160 Tbilisi, Georgia}

\begin{abstract}

A novel model of incompressible magnetohydrodynamic turbulence in
the presence of a strong external magnetic field is proposed for
explanation of recent numerical results. According to the proposed
model, in the presence of the strong external magnetic field,
incompressible magnetohydrodynamic turbulence becomes nonlocal in
the sense that low frequency modes cause decorrelation of
interacting high frequency modes from the inertial interval. It is
shown that the obtained nonlocal spectrum of the inertial range of
incompressible magnetohydrodynamic turbulence represents an
anisotropic analogue of Kraichnan's nonlocal spectrum of
hydrodynamic turbulence. Based on the analysis performed in the
framework of the weak coupling approximation, which represents one
of the equivalent formulations of the direct interaction
approximation, it is shown that incompressible magnetohydrodynamic
turbulence could be both local and nonlocal and therefore
anisotropic analogues of both the Kolmogorov and Kraichnan spectra
are realizable in incompressible magnetohydrodynamic turbulence.

\end{abstract}

\pacs{52.35.Ra, 47.27.Gs, 47.27.Eq}

\maketitle

\section{Introduction} \label{sec:1}

Although  magnetohydrodynamic (MHD) turbulence has been extensively
studied for the last 40 years
\cite{I63,Kr65a,SMM83,MZ89,BSD96,GS95,GS97,CV00,MG01,CLV02,MBG03,HBD04,BS04,MG05,B05,OM05,BL06,LGS06}
many physical aspects of the problem still remain unclear (for
recent reviews see Refs. \cite{B03,V04,SC05,K05}). The first model
of incompressible MHD turbulence was proposed by Iroshnikov
\cite{I63} and Kraichnan \cite{Kr65a}. The Iroshnikov-Kraichnan (IK)
model of MHD turbulence is based on the so-called Alfv\'en effect
\cite{B03} - nonlinear interaction is possible only among Alfv\'en
waves propagating in opposite directions along the mean magnetic
field. Therefore, IK model assumes that the energy cascade in MHD
turbulence occurs as a result of collisions between oppositely
propagating Alfv\'en waves. Consider isotropic excitation of
Alfv\'en waves on some outer scale $l_0$ with a characteristic
velocity $v_0 \ll V_A$, where $V_A$ is the Alfv\'en velocity. IK
model assumes that the energy transfer is local and isotropic in the
wave number space. The characteristic time scale of the Alfv\'en
wave collision is $\tau_{ac}^{IK} \sim (V_A k)^{-1}$, where $k$ is
the wave number. Using the governing equations of incompressible MHD
it can be shown that during one collision distortion of each wave
packet $\delta v_l$ is of order
\begin{equation}
\frac{\delta v_l}{v_l} \sim \frac{v_l}{V_A} \ll 1. \label{eq:0.1}
\end{equation}
Because these perturbations are summed with random phases $N \sim
(v_l/ \delta v_l)^2 \sim (V_A / v_l)^2$ collisions are necessary to
achieve the distortion of order unity. Therefore for the energy
cascade time $\tau_{cas}^{IK}$ we have
\begin{equation}
\tau_{cas}^{IK} \sim \frac{1}{k v_l} \frac{V_A}{v_l}. \label{eq:0.2}
\end{equation}
Taking into account the relations $\varepsilon \sim
v_l^2/\tau_{cas}$, $v_l^2 \sim k E_k$, where $\varepsilon $ is the
energy cascade rate and $E_k$ is the one dimensional energy
spectrum, we obtain
\begin{equation}
\frac{v_l}{v_0} \sim \left( \frac{l}{l_0} \right)^{1/4},~~~E_k^{IK}
\sim (\varepsilon V_A)^{1/2} k^{-3/2}, \label{eq:0.3}
\end{equation}
which represents IK spectrum of incompressible MHD turbulence.

IK model of the MHD turbulence is isotropic. However, presence of a
mean magnetic field has a strong effect on the turbulence
properties, in contrast to a mean flow in hydrodynamic turbulence,
which can be eliminated by a Galilean transformation. The anisotropy
of MHD turbulence had been seen in various numerical simulations
\cite{SMM83,BSD96,MG01,MBG03,HBD04}. A theory of anisotropic MHD
turbulence was proposed by Goldreich and Sridhar (GS) \cite{GS95}.
GS model implies that the dynamics of turbulence is dominated by the
perpendicular cascade (i.e., by the cascade in the ${\bf q}$-space,
where ${\bf q}$ is the component of the wave number vector
perpendicular to the mean magnetic field), whereas the parallel size
of turbulent 'eddies' (wave packets) is determined by critical
balance condition which implies that the characteristic time scale
of wave collision $(pV_A)^{-1}$, where $p$ is parallel wave number,
is equal to the characteristic time scale the energy cascade
$\tau_{cas}$. GS model also assumes that the energy cascade is local
in the wave number space and therefore $\tau_{cas}^{GS} \sim (v_l
q)^{-1}$. This yields (we suppose that turbulence is generated at
perpendicular scales $l_{q_0} \sim 1/q_0$ with a characteristic
velocity $v_0$)
\begin{equation}
\frac{v_l}{v_0} \sim \left( \frac{l_q}{l_{q_0}}
\right)^{1/3},~~~E_k^{GS} \sim \varepsilon^{2/3} q^{-5/3},
\label{eq:0.4}
\end{equation}
which represents the anisotropic analogue of the Kolmogorov spectrum
of isotropic hydrodynamic turbulence. The anisotropy of fluctuations
is determined by the critical balance principle, which yields $p
\sim q^{\mu}$, with $\mu =2/3$, and therefore as the cascade
proceeds to larger $q$, the eddies become more elongated along the
direction of the mean magnetic field.

however, GS model does not totally agree with the results of the
recent numerical simulations \cite{MG01,MBG03,HBD04,MG05}. It has
been found \cite{MG01} that in the presence of the strong external
magnetic field incompressible MHD turbulence is strongly anisotropic
with $\mu \approx 1/2$ and has the one dimensional spectrum similar
to IK one $E_k \sim q^{-3/2}$. As it was shown in Ref. \cite{MBG03}
the anisotropic spectrum of incompressible MHD turbulence strongly
depends on the strength of the external magnetic field. It was found
that the perpendicular scaling of fluctuations changed from GS form
to the form found in Ref. \cite{MG01} as the external field was
increased from $\gamma \equiv B_0^2/\rho v_{0}^2 \ll 1$, where $B_0$
is the external magnetic field and $\rho$ is the density, to the
limit $\gamma \gg 1$.

For resolution of this inconsistency it has been noted \cite{B05}
that if in the presence of the strong external magnetic field
nonlinear interaction of the counter propagating fluctuations is
reduced by the factor $v_l/V_A$, so that the perpendicular cascade
timescale becomes
\begin{equation}
\tau_{cas}^{AIK} \sim \frac{1}{k v_{l_q}} \frac{V_A}{v_{l_q}},
\label{eq:0.5}
\end{equation}
then for the one dimensional spectrum one obtains $E_q^{AIK} \sim
(\varepsilon V_A)^{1/2} q^{-3/2}$. This model of MHD turbulence
represents an anisotropic analogue of IK model. The characteristic
parallel scale of the turbulent eddies is determined by the
condition similar to GS critical balance condition $1/p \sim V_A
\tau_{cas}$, which yields $p \sim q^{1/2}$.

Although anisotropic IK model properly represents scaling indices of
incompressible MHD turbulence in the presence of the strong external
magnetic field, it still contradicts with some other features of the
turbulence observed in the numerical simulations. Indeed, according
to Eq. (\ref{eq:0.5}), the ratio of the cascade timescale to the
eddy turnover time $\tau_{to} \sim (k v_l)^{-1}$ is
$\tau_{cas}/\tau_{to} \sim v_l/V_A$. For the parameters used in Ref.
\cite{MG01} the ratio $v_l/V_A \sim 10^{-2}-10^{-3}$, whereas the
ratio $\tau_{cas}/\tau_{to} \sim 1$ (see Sec. 5.6.3 of Ref
\cite{MG01}). Therefore, anisotropic IK model gives strongly
understated estimate for the characteristic cascade rate.

In this paper we present a novel model of incompressible MHD
turbulence. The main assumption of the model is the nonlocal
character of MHD turbulence in the presence of a strong external
magnetic field. The model implies that low frequency modes cause
decorrelation of interacting high frequency modes from the inertial
interval. Proposed model represents an anisotropic analogue of
Kraichnan's nonlocal spectrum of hydrodynamic turbulence \cite{K59}
and properly reproduces main features of incompressible MHD
turbulence observed in the numerical simulations.

Our study of incompressible MHD turbulence is performed in the
framework of Kadomtsev's weak coupling approximation (WCA)
\cite{K65}, which represents one of the equivalent formulations of
Kraichnan's direct interaction approximation (DIA) \cite{K59}. Our
analysis does not allow us to prove that in the presence of a strong
external magnetic field incompressible MHD turbulence becomes
nonlocal. However, performed analysis shows that for the systems
with more then one degrees of freedom the concept of adiabatic and
resonant interactions does not hold and consequently, in contrast to
isotropic hydrodynamic turbulence which is local, incompressible MHD
turbulence could be both local and nonlocal. Obtained local solution
reproduces GS model of MHD turbulence, whereas obtained nonlocal
solution represents an anisotropic analogue of Kraichnan's nonlocal
spectrum of hydrodynamic turbulence.

The paper is organized as follows. The heuristics of the proposed
model is presented in Sec. \ref{sec:2}. Physics of the decorrelation
mechanism is discussed in Sec. \ref{sec:3}. The WCA equations of
incompressible MHD turbulence are derived in Sec. \ref{sec:4}.
Analysis of the WCA equations of isotropic hydrodynamic turbulence
is given in Sec. \ref{sec:5}. Local as well as nonlocal solutions of
the WCA equations of incompressible MHD turbulence are obtained in
Sec. \ref{sec:6}. Locality of incompressible MHD turbulence is
discussed in Sec. \ref{sec:7}. Conclusions are given in Sec.
\ref{sec:8}.

\section{Heuristics of the model} \label{sec:2}

Kraichnan obtained nonlocal spectrum of hydrodynamic turbulence as a
possible inertial range solution of the Eulerian DIA equations
\cite{K59}. Kraichnan's nonlocal spectrum of isotropic hydrodynamic
turbulence can be derived based on the following qualitative
arguments \cite{Kr}: assume that the autocorrelation time scale is
determined by advection of a small-scale eddies (from the inertial
range) of the size $k^{-1}$ through a distance of the order of its
own size by a macroscopic flow of rms velocity  which is determined
by the characteristic velocity of energy containing eddies $v_0$,
i.e., assume that $\tau_{ac}^{Kr} \sim 1/(v_0 k)$. Then the energy
transfer rate reduces by the factor $\tau_{ac}^{Kr}/\tau_{to}$
compared to the Kolmogorov theory (here $\tau_{to} \sim 1/(v_l k)$
is the characteristic eddy turnover timescale) and therefore the
characteristic timescale of the turbulent cascade becomes
\begin{equation}
\tau_{cas}^{Kr} \sim \frac{1}{k v_l} \frac{v_0}{v_l}. \label{eq:0.6}
\end{equation}
For the velocity fluctuations and the one dimensional energy
spectrum one obtains $v_l \sim l^{1/4}$ and $E_k^{K} \sim
(\varepsilon v_0)^{1/2} k^{-3/2}$, respectively. This represents
Kraichnan's nonlocal spectrum of isotropic hydrodynamic turbulence
\cite{K59}.

As it was proposed by Richardson \cite{R} and confirmed by various
experiments (see, e.g., Refs. \cite{Le,H} and references therein)
the energy cascade in isotropic hydrodynamic turbulence is local and
therefore the Kolmogorov spectrum is the only one that is realized
in the inertial range of isotropic hydrodynamic turbulence.

We propose that in contrast to hydrodynamic turbulence,
incompressible MHD turbulence is  non-local in the presence of the
strong ($\gamma \gg 1$) external magnetic field, i.e., an influence
of low frequency modes leads to decorrelation of interacting high
frequency modes from the inertial interval. Physics of the
decorreelation mechanism is discussed in the next section. We argue
that $E_q \sim q^{-3/2}$ one dimensional spectrum observed in the
numerical simulations represents the anisotropic analogue of the
Kraichnan nonlocal spectrum. In particular, similar to GS model,
presented model implies that incompressible energy transfer is
strongly anisotropic and dominated by perpendicular cascade. The
later cascade is assumed to be nonlocal, and similar to Eq.
(\ref{eq:0.6}) we have for the characteristic cascade timescale
\begin{equation}
\tau_{cas} \sim \frac{1}{k v_{l_q}} \frac{v_0}{v_{l_q}}.
\label{eq:0.7}
\end{equation}
therefore for the velocity fluctuations and the one dimensional
energy spectrum our model implies
\begin{equation}
\frac{v_{l_q}}{v_0} \sim \left( \frac{l_q}{l_{q_0}}
\right)^{1/4},~~~E_q \sim (\varepsilon v_0)^{1/2} q^{-3/2}.
\label{eq:0.8}
\end{equation}
The relation between the parallel and perpendicular sizes of
turbulent eddies is determined based on the following arguments
\cite{GS95}: the lifetime of a wave packet is of order $\tau_{cas}$
and therefore for the correlation lengthscale along the field line
we have
\begin{equation}
l_p \sim 1/p \sim \tau_{cas} V_A. \label{eq:0.9}
\end{equation}
Using Eq. (\ref{eq:0.7}) this yields
\begin{equation}
p \sim \frac{v_0}{V_A} q_0^{1/2} q^{1/2}. \label{eq:0.10}
\end{equation}

Presented model properly reproduces main characteristics of
incompressible MHD turbulence observed in the numerical simulations
\cite{MG01,MBG03}.

Explicit dependence of the cascade rates given by Eqs.
(\ref{eq:0.6}) and (\ref{eq:0.7}) on the characteristic velocity of
low frequency energy containing modes represents important feature
of Krichnan's nonlocal spectrum as well as considered model and
underlines the nonlocal character of the turbulence.

\section{Physics of the decorrelation mechanism} \label{sec:3}

The governing equations of incompressible magnetohydrodynamics in
the Elsasser variables are
\begin{equation}
\partial_t {\bf U} = - ({\bf W} \cdot {\bf \nabla})
{\bf U} - {\bf \nabla}P + {\bar \nu} {\bf \nabla}^2 {\bf U} + {\bar
\nu}_m {\bf \nabla}^2 {\bf W}, \label{eq:1.1}
\end{equation}
\begin{equation}
\partial_t {\bf W} = - ({\bf U} \cdot {\bf \nabla})
{\bf W}- {\bf \nabla}P + {\bar \nu}_m {\bf \nabla}^2 {\bf U} + {\bar
\nu} {\bf \nabla}^2 {\bf W}, \label{eq:1.2}
\end{equation}
\begin{equation}
{\bf \nabla}\cdot {\bf W}={\bf \nabla}\cdot {\bf U}=0,
\label{eq:1.3}
\end{equation}
where ${\bf U}={\bf v}+{\bf b}$ and  ${\bf W}={\bf v}-{\bf b}$ are
the Elsasser variables, ${\bf v}$ is the velocity field, ${\bf
b}\equiv {\bf B}/\sqrt{4\pi \rho}$ is the magnetic field in velocity
units, $P$ is the total (hydrodynamic plus magnetic) pressure
normalized by the density, $\partial_t\equiv \partial/\partial t$,
${\bar \nu} \equiv (\nu + \nu_m)/2$, ${\bar \nu}_m \equiv (\nu -
\nu_m)/2$, $\nu$ is the kinematic viscosity and $\nu_m$ is the
magnetic diffusivity.

Consider the packet of Alfv\'en waves propagating along the uniform
magnetic field ${\bf \bar b}_0 \parallel z$, with the characteristic
magnetic field perturbation ${\bf b}_0 \perp z$, the velocity
perturbation ${\bf v}_0 = - {\bf b}_0$ and the characteristic
parallel and perpendicular wave numbers $p_0$ and $q_0$,
respectively. On this background consider a high frequency wave
packet propagating in the opposite direction with the characteristic
parallel and perpendicular wave numbers $p_u\gg p_0$ and $q_u \gg
q_0$, respectively. Denote the Elsasser variable associated with the
high frequency wave packet by ${\bf u}_u$. Assuming ${\bf b}_0$ as
approximately constant and dropping dissipation terms Eq.
(\ref{eq:1.1}) yields
\begin{equation}
(\partial_t - {\bf \bar b}_0 \cdot \nabla - {\bf b}_0 \cdot \nabla +
{\bf v}_0 \cdot \nabla) {\bf u}_u \approx 0, \label{eq:1.2a}
\end{equation}
Noting that ${\bf v}_0 = - {\bf b}_0$, this equation gives for
characteristic frequencies of the high frequency packet
\begin{equation}
\omega_u \approx -{\bar b}_0 p_u + 2 v_0 q_u. \label{eq:1.2c}
\end{equation}
If we consider dynamics of a high frequency Alfv\'en wave packet
propagating parallel to $z$-axis (with the characteristic parallel
and perpendicular wave numbers $p_w \sim p_u \sim p$ and $q_w \sim
q_u \sim q$ respectively), then similar consideration yields for the
frequency
\begin{equation}
\omega_w \approx {\bar b}_0 p_w. \label{eq:1.2d}
\end{equation}
Therefore, presence of the low frequency wave packet do not
influence propagation of the high frequency packet along $z$-axis
(Alfv\'en effect) whereas the packet propagating in the opposite
direction is moved with the velocity $2v_0$ in the direction
perpendicular to $z$-axis. This circumstance represents the bases
for understanding, why incompressible MHD turbulence becomes
nonlocal in the presence of the strong external magnetic field.
Indeed, consider two interacting packets of high frequency alfv\'en
waves moving in opposite directions on the same background as above
(i.e., in the presence of strong external magnetic field and the low
frequency Alfv\'en wave packet). The perpendicular length scale of
the packets is of order $1/q$ and consequently, due to the described
above action of the low frequency packet, the interacting packets
would be pulled apart during the time scale $1/(q v_0)$. Therefore,
the characteristic timescale of the unit act of interaction shortens
compared to the Kolmogorov autocorrelation timescale $1/(q v_l)$. As
it was shown in the previous section this automatically implies that
the turbulence would have the spectral characteristics given by Eqs.
(\ref{eq:0.7})-(\ref{eq:0.10}).

In the consideration above we implicitly assumed that the low
frequency packet do not contribute to the mean magnetic field which
acts on the interacting high frequency wave packets (i.e., it has
been assumed that the anisotropic high frequency wave packets are
formed along the external magnetic field ${\bf \bar b}_0$). If this
is not the case then the situation changes entirely. Indeed, assume
that the low frequency wave packet contributes to the mean field
acting on high frequency packets, i.e., assume that the high
frequency packets are formed along the axis ${\bf \bar b}_0+{\bf
b}_0$ instead of ${\bf \bar b}_0$. In this case, instead of Eqs.
(\ref{eq:1.2c})-(\ref{eq:1.2d}) similar analysis yield
\begin{equation}
\omega_u \approx -({\bar b}_0 + {b}_0) p_u + {\bf v}_0 \cdot {\bf
q}_u, \label{eq:1.2e}
\end{equation}
\begin{equation}
\omega_w \approx ({\bar b}_0 + {b}_0) p_w + {\bf v}_0 \cdot {\bf
q}_w. \label{eq:1.2f}
\end{equation}
The second terms in Eqs. (\ref{eq:1.2e})-(\ref{eq:1.2f}) describes
the Doppler shift caused by the velocity field of the low frequency
wave packet and could be removed by corresponding Galilean
transformation. Consequently, in this case the only effect caused by
the presence of the low frequency modes is the change of the mean
field, and they do not cause the decorrelation of high frequency
modes. This arguments lead us to the conclusion that {\it only the
low frequency modes which do not contribute to the mean field cause
decorrelation} of interacting high frequency packets.

Consequently, presented model implies that in the presence of the
strong external magnetic field $\bar b_0 \gg b_0$ the low frequency
energy containing modes do not contribute to the mean field which
acts on the high frequency fluctuations in the inertial interval. As
a result, decorrelation mechanism described above causes formation
of the anisotropic analogue of the Kraichnan nonlocal spectrum. On
the other hand, in the absence of the strong external magnetic
field, the mean field is formed by the low frequency (energy
containing) modes, the nonlocal decorrelation mechanism is not at
work, turbulence is local and therefore is described by GS model.

\section{The WCA equations for incompressible MHD turbulence} \label{sec:4}

Consider incompressible MHD turbulence in the presence of the
constant magnetic field ${\bf B}_0$ directed along $z$ axis.
Presenting the Elsasser variables as a sum of the mean and
fluctuating parts ${\bf W}={\bf W}_0+{\bf W}_1$,~${\bf U}={\bf
U}_0+{\bf U}_1$, where  ${\bf U}_0={\bf V}_A$ and ${\bf W}_0=-{\bf
V}_A$, and ${\bf V}_A \equiv {\bf B_0}/\sqrt{4\pi \rho}$ is the
Alfv\'en velocity, and dropping dissipation terms Eqs.
(\ref{eq:1.1})-(\ref{eq:1.3}) yield
\begin{equation}
\partial_t {\bf U}_1 - {\bf V}_A \partial_z {\bf U}_1 = - ({\bf W}_1 \cdot {\bf \nabla})
{\bf U}_1 - {\bf \nabla}p \label{eq:2.1}
\end{equation}
\begin{equation}
\partial_t {\bf W}_1 + {\bf V}_A \partial_z {\bf W}_1 = - ({\bf U}_1 \cdot {\bf \nabla})
{\bf W}_1- {\bf \nabla}p \label{eq:2.2}
\end{equation}

Performing the Fourier transform defined as
\begin{equation}
{\bf u}_{{\bf k}, \omega}= \frac{1}{(2\pi)^4} \int \exp(i\omega
t-i{\bf k \cdot x}) {\bf U}_1({\bf x},t) d^3{\bf x} dt,
 \label{eq:3}
\end{equation}
and eliminating pressure terms we obtain
\begin{equation}
(\omega+\omega_{\bf k}) {\bf u}_{\bf \bar k}= \int [{\bf u_1}-{\bf
\hat k}({\bf \hat k} \cdot {\bf u_1})]({\bf k} \cdot {\bf w_2})
d{\cal F}_{1,2}^k,
 \label{eq:4.1}
\end{equation}
\begin{equation}
(\omega-\omega_{\bf k}) {\bf w}_{\bf \bar k}= \int [{\bf w_1}-{\bf
\hat k}({\bf \hat k} \cdot {\bf w_1})]({\bf k} \cdot {\bf u_2}) d
{\cal F}_{1,2}^k,
 \label{eq:4.2}
\end{equation}
where ${\bf \bar k} \equiv ({\bf k},\omega)$, the caret denotes the
unit vector, ${\bf u}_1$ denotes ${\bf u_{{\bar k}_1}}$,~
$\omega_{\bf k}=V_A k_z$ is the frequency of the Alfv\'en wave, $d
{\cal F}_{1,2}^k \equiv d^4 {\bf \bar k}_1 d^4 {\bf \bar k}_2
\delta_{{\bf \bar k}-{\bf \bar k}_1 - {\bf \bar k}_2}$, and
$\delta_{{\bf \bar k}-{\bf \bar k}_1 - {\bf \bar k}_2} \equiv
\delta({\bf \bar k}-{\bf \bar k}_1 - {\bf \bar k}_2)$ is the Dirac
delta function.

Incompressible MHD turbulence is governed by interaction of shear
Alfv\'en waves, whereas pseudo Alfv\'en waves play a passive role
\cite{GS95,GS97}. Therefore, in the presented paper we consider the
shear Alfv\'enic turbulence. Defining the unit polarization vector
of the shear Alfv\'en waves ${\bf \hat e_{k}}= {\bf \hat k} \times
{\bf z}$, and introducing the amplitudes of the shear Alfv\'en waves
as
\begin{equation}
{\bf w_{\bar k}}=i \phi_{\bf \bar k} {\bf \hat e_{k}},~~~~{\bf
u_{\bar k}}=i \psi_{\bf \bar k} {\bf \hat e_{k}}, \label{eq:5}
\end{equation}
Eqs. (\ref{eq:4.1})-(\ref{eq:4.2}) reduce to the following equations
\begin{equation}
(\omega-\omega_{\bf k}) \phi_{\bf \bar k}= \int_{-\infty}^{\infty}
T_{1,2} \phi_{1} \psi_{2} d{\cal F}_{1,2}^k,
 \label{eq:6.1}
\end{equation}
\begin{equation}
(\omega+\omega_{\bf k}) \psi_{\bf \bar k}= \int_{-\infty}^{\infty}
T_{1,2} \psi_{1} \phi_{2} d{\cal F}_{1,2}^k, \label{eq:6.2}
\end{equation}
where $T_{1,2} \equiv i({\bf \hat e}_{\bf k} \cdot {\bf \hat e}_{\bf
k_1})({\bf k} \cdot {\bf \hat e}_{\bf k_2})$ is the matrix element
of interaction.

To achieve any progress in analysis of Eqs.
(\ref{eq:6.1})-(\ref{eq:6.2}) some closure scheme should be used. In
Ref. \cite{GS95} the eddy damped quasi normal Markovian (EDQNM)
approximation was used for the analysis of Eqs.
(\ref{eq:6.1})-(\ref{eq:6.2}). In the framework of this
approximation the linear damping term is added to the equation for
the third order moments of the turbulent fields and afterwards some
assumptions are made regarding the eddy damping rate. But an
assumption of some {\it a priori} form of the eddy damping rate
automatically fixes the property of the locality of turbulence - is
it assumed to be local or not. Note that locality of the turbulence
is one of the main assumptions of GS model \cite{GS95}. Here we use
the WCA (DIA) for the study of locality of the incompressible MHD
turbulence.

In the framework of the DIA \cite{K59} one differentiates direct
nonlinear interactions among three Fourier modes with the complanar
wave vectors (${\bf k}+{\bf k}_1+{\bf k}_2=0$) and indirect
interactions when the modes are interacting by means of other
Fourier modes. The idea of the closure of equations for the
turbulent fields is based on the assumption that the influence of
the indirect interactions on the turbulence dynamics can be
neglected in comparison with the direct interactions.

We use the standard WCA technique \cite{K65} for the derivation of
the WCA equations of incompressible MHD turbulence. One of the main
nonlinear effects described by the nonlinear interaction terms on
the right hand sides of Eqs. (\ref{eq:6.1})-(\ref{eq:6.2}) is the
nonlinear decay of the mode $\phi_{\bf \bar k}$. The intensity of
this process is proportional to the amplitude of the mode. The WCA
implies the isolation of this part of the nonlinear interaction. For
this purposes one should add to both sides of Eqs.
(\ref{eq:6.1})-(\ref{eq:6.2}) the terms $i \zeta_{\bf \bar k}^+
\phi_{\bf {\bar k}}$ and $i\zeta_{\bf \bar k}^- \psi_{\bf {\bar
k}}$, respectively. Multiplying obtained equations respectively by
$\phi_{\bf {\bar k}^\prime}^\ast$ and $\psi_{\bf {\bar
k}^\prime}^\ast$ (here and hereafter asterisk denotes the complex
conjugated value), ensemble averaging, considering different wave
modes as statistically independent and introducing the following
notations
\begin{equation}
\langle \phi_{\bf {\bar k}} \phi_{\bf {\bar k}^\prime}^\ast \rangle
=I_{\bf {\bar k}}^+ \delta_{\bf {\bar k}-\bf {\bar k}^\prime},~~~
\langle \psi_{\bf {\bar k}} \psi_{\bf {\bar k}^\prime}^\ast \rangle
=I_{\bf {\bar k}}^- \delta_{\bf {\bar k}-\bf {\bar k}^\prime},
 \label{eq:7}
\end{equation}
we obtain
\begin{eqnarray}
(\omega - \omega_{\bf k} + i \zeta_{\bf \bar k}^+) I_{\bf {\bar
k}}^+ \delta_{\bf {\bar k}-\bf {\bar k}^\prime} = i \zeta_{\bf \bar
k}^+ I_{\bf {\bar k}}^+ \delta_{\bf {\bar k}-\bf {\bar k}^\prime}+
\nonumber \\ \int_{-\infty}^\infty T_{1,2} \langle \phi_{1} \psi_{2}
\phi_{\bf \bar{k}^\prime}^\ast \rangle d{\cal F}_{1,2}^k,
 \label{eq:8.1}
\end{eqnarray}
\begin{eqnarray}
(\omega + \omega_{\bf k} + i \zeta_{\bf \bar k}^-) I_{\bf {\bar
k}}^- \delta_{\bf {\bar k}-\bf {\bar k}^\prime} = i \zeta_{\bf \bar
k}^- I_{\bf {\bar k}}^- \delta_{\bf {\bar k}-\bf {\bar k}^\prime}+
\nonumber \\ \int_{-\infty}^\infty T_{1,2} \langle \psi_{1} \phi_{2}
\psi_{\bf \bar{k}^\prime}^\ast \rangle d{\cal F}_{1,2}^k.
 \label{eq:8.2}
\end{eqnarray}

The second terms on the right hand sides of Eqs.
(\ref{eq:8.1})-(\ref{eq:8.2}) contain contributions from both direct
and indirect nonlinear interactions. Following the procedure
developed in Ref. \cite{K65} for the elimination of the contribution
of the indirect nonlinear interactions, let us represent the
turbulent fields as
\begin{equation}
\phi_{\bf {\bar k}}=\phi_{\bf {\bar k}}^{(0)}+\phi_{\bf {\bar
k}}^{(1)},~~~~\phi_{\bf {\bar k}}^{(0)} \ll \phi_{\bf {\bar
k}}^{(1)}, \label{eq:9.1}
\end{equation}
\begin{equation}
\psi_{\bf {\bar k}}=\psi_{\bf {\bar k}}^{(0)}+\psi_{\bf {\bar
k}}^{(1)},~~~~\psi_{\bf {\bar k}}^{(0)} \ll \psi_{\bf {\bar
k}}^{(1)}, \label{eq:9.2}
\end{equation}
where in the zeroth approximation the fluctuations are assumed to be
uncorrelated. Then the third order moments on right hand sides of
Eqs. (\ref{eq:8.1})-(\ref{eq:8.2}) vanish in the zeroth order
approximation. In the first order approximation we have
\begin{eqnarray}
\langle \phi_1 \psi_2 \phi_{\bf \bar k}^\ast \rangle = \langle
\phi_1^{(1)} \psi_2^{(0)} \phi_{\bf \bar k}^{\ast {(0)}} \rangle +
\langle \phi_1^{(0)} \psi_2^{(1)} \phi_{\bf \bar k}^{\ast {(0)}}
\rangle+ \nonumber \\ \langle \phi_1^{(0)} \psi_2^{(0)} \phi_{\bf
\bar k}^{\ast {(1)}} \rangle. \label{eq:10}
\end{eqnarray}
For $\phi_{\bf \bar k}^{(1)}$ Eq. (\ref{eq:6.1}) gives
\begin{equation}
(\omega-\omega_{\bf k}) \phi_{\bf \bar k}^{(1)}=
\int_{-\infty}^\infty T_{1,2} \phi_{1}^{(0)} \psi_{2}^{(0)} d{\cal
F}_{1,2}^k.
 \label{eq:11}
\end{equation}
To isolate the contribution of only direct nonlinear interactions we
note, that the duration of the unit act of the nonlinear interaction
is $t_\zeta \sim 1/\zeta_{\bf {\bar k}}^+$. Therefore, to hold
contribution of only direct nonlinear interactions we should replace
$(\omega - \omega_{\bf k})$ by $(\omega - \omega_{\bf k} +
i\zeta_{\bf \bar k}^+)$ in Eq. (\ref{eq:11}). This yields
\begin{equation}
\phi_{\bf \bar k}^{(1)}= \frac{1}{(\omega-\omega_{\bf k} +
i\zeta_{\bf \bar k}^+)} \int_{-\infty}^\infty T_{1,2} \phi_{1}^{(0)}
\psi_{2}^{(0)} d{\cal F}_{1,2}^k.
 \label{eq:12}
\end{equation}
Similarly for $\psi_{\bf \bar k}^{(1)}$ we obtain
\begin{equation}
\psi_{\bf \bar k}^{(1)}= \frac{1}{(\omega + \omega_{\bf k} +
i\zeta_{\bf \bar k}^-)} \int_{-\infty}^\infty T_{1,2} \psi_{1}^{(0)}
\phi_{2}^{(0)} d{\cal F}_{1,2}^k.
 \label{eq:13}
\end{equation}

Substituting Eqs. (\ref{eq:10}), (\ref{eq:12}) and (\ref{eq:13})
into Eqs. (\ref{eq:8.1})-(\ref{eq:8.2}), considering the fields
$\phi_{\bf \bar k}$ and $\psi_{\bf \bar k}$ as uncorrelated (i.e.,
assuming $\langle \phi_1 \psi_2 \rangle \approx 0$~~\cite{GS95},
which physically implies that we consider MHD turbulence with zero
residual energy \cite{B03,MG05} or equivalently with equal kinetic
and magnetic energies), using for the forth order moments the
Gaussian relations
\begin{eqnarray}
&& \langle \phi_3 \psi_4 \psi_2 \phi_{\bf \bar k}^\ast \rangle =
\langle \phi_3 \phi_{\bf \bar k} \rangle \langle \psi_4 \psi_2
\rangle = I_{\bf \bar k}^+ I_2^- \delta_{{\bf \bar k} - {\bf \bar
k}_3} \delta_{{\bf \bar k}_2 + {\bf \bar k}_4}
\nonumber \\
&& \langle \phi_1 \psi_3 \phi_4 \phi_{\bf \bar k}^\ast \rangle = 0
\\ && \langle \phi_1 \psi_2 \phi_3^\ast \psi_4^\ast
\rangle = \langle \phi_1 \phi_3^\ast \rangle \langle \psi_2
\psi_4^\ast \rangle = I_1^+ I_2^- \delta_{{\bf \bar k}_1 - {\bf \bar
k}_3} \delta_{{\bf \bar k}_2 - {\bf \bar k}_4} \nonumber,
\label{eq:14}
\end{eqnarray}
and taking into account that according to the definition of
$\zeta_{\bf \bar k}^\pm$, the right hand sides of  Eqs.
(\ref{eq:8.1})-(\ref{eq:8.2}) should not contain the terms
proportional to $I_{\bf \bar k}^\pm$ we finally arrive at the
following equations
\begin{equation}
i\zeta_{\bf \bar k}^+ = - \int_{-\infty}^\infty \frac{T_{1,2}
T_{{\bf k},-2}}{\omega_1 - \omega_{{\bf k}_1} + i\zeta_1^+} I_2^-
d{\cal F}_{1,2}^k,
 \label{eq:15.1}
\end{equation}
\begin{equation}
i\zeta_{\bf \bar k}^- = - \int_{-\infty}^\infty \frac{T_{1,2}
T_{{\bf k},-2}}{\omega_1 + \omega_{{\bf k}_1} + i\zeta_1^-} I_2^+
d{\cal F}_{1,2}^k,
 \label{eq:15.10}
\end{equation}
\begin{equation}
|(\omega-\omega_{\bf k} + i\zeta_{\bf \bar k}^+)|^2 I_{\bf \bar k}^+
= \int_{-\infty}^\infty |T_{1,2}|^2 I_1^+ I_2^- d{\cal F}_{1,2}^k,
 \label{eq:15.20}
\end{equation}
\begin{equation}
|(\omega+\omega_{\bf k} + i\zeta_{\bf \bar k}^-)|^2 I_{\bf \bar k}^-
= \int_{-\infty}^\infty |T_{1,2}|^2 I_1^- I_2^+ d{\cal F}_{1,2}^k.
 \label{eq:15.2}
\end{equation}

Eqs. (\ref{eq:15.1})-(\ref{eq:15.2}) represent the WCA (equivalently
the Eulerian DIA) equations for incompressible MHD turbulence.
Noting that $T_{1,2}=T_{{\bf k},-2}^\ast$ and defining
\begin{equation}
\Gamma_{\bf \bar k}^\pm =  \frac{i}{\omega \mp \omega_{\bf k} +
i\zeta_{\bf \bar k}^\pm},
 \label{eq:16}
\end{equation}
after straightforward manipulations Eqs.
(\ref{eq:15.1})-(\ref{eq:15.2}) can be rewritten as
\begin{equation}
-i(\omega-\omega_{\bf k}) \Gamma_{\bf \bar k}^+ = 1 - \Gamma_{\bf
\bar k}^+ \int_{-\infty}^\infty |T_{1,2}|^2 \Gamma_1^+ I_2^- d{\cal
F}_{1,2}^k,
 \label{eq:17.1}
\end{equation}
\begin{equation}
-i(\omega + \omega_{\bf k}) \Gamma_{\bf \bar k}^- = 1 - \Gamma_{\bf
\bar k}^- \int_{-\infty}^\infty |T_{1,2}|^2 \Gamma_1^- I_2^+ d{\cal
F}_{1,2}^k,
 \label{eq:17.10}
\end{equation}
\begin{eqnarray}
-i(\omega-\omega_{\bf k}) I_{\bf \bar k}^+ = {\Gamma_{\bf \bar
k}^+}^\ast \int_{-\infty}^\infty |T_{1,2}|^2 I_1^+ I_2^- d{\cal
F}_{1,2}^k - \nonumber \\ I_{\bf \bar k}^+ \int_{-\infty}^\infty
|T_{1,2}|^2 \Gamma_1^+ I_2^- d{\cal F}_{1,2}^k, \label{eq:17.20}
\end{eqnarray}
\begin{eqnarray}
-i(\omega + \omega_{\bf k}) I_{\bf \bar k}^- = {\Gamma_{\bf \bar
k}^-}^\ast \int_{-\infty}^\infty |T_{1,2}|^2 I_1^- I_2^+ d{\cal
F}_{1,2}^k - \nonumber \\ I_{\bf \bar k}^- \int_{-\infty}^\infty
|T_{1,2}|^2 \Gamma_1^- I_2^+ d{\cal F}_{1,2}^k,
 \label{eq:17.2}
\end{eqnarray}

The aim of the further analysis is the study of Eqs.
(\ref{eq:15.1})-(\ref{eq:15.2}), or equivalently,
(\ref{eq:17.1})-(\ref{eq:17.2}) and similar equations for isotropic
hydrodynamic turbulence. But before performing this analysis we
shortly discuss several topics related to the nature of the obtained
equations.

\subsection{The equivalence of the WCA and the DIA}

As it was mentioned above, the WCA is one of the equivalent forms of
the DIA. Essentially the WCA is the DIA formulated in the frequency
domain instead of the time domain \cite{Kr,DP85}. To perform the DIA
analysis of incompressible MHD turbulence one should start from Eqs.
(\ref{eq:6.1})-(\ref{eq:6.2}) in the time domain (Eqs. (15) of Ref.
\cite{GS95}), add random forcing terms and define the Green
functions $G^+({\bf k},t-t^\prime)$ and $G^-({\bf k},t-t^\prime)$ of
the obtained equations. Then using standard assumptions of this
approach (see, e.g., \cite{Kr} and references therein) for
separation of the direct and indirect interactions one should obtain
the closed set of equations for functions $Q^+({\bf k},t,t^\prime)$
and $Q^-({\bf k},t,t^\prime)$ defined as
\begin{equation}
\langle \tilde{\phi}_{\bf k}(t)  \tilde{\phi}_{\bf
{k^\prime}}(t^\prime) \rangle \equiv Q^+({\bf k},t,t^\prime)
\delta_{{\bf k} - {\bf k^\prime}},
 \label{eq:18.1}
\end{equation}
\begin{equation}
\langle \tilde{\psi}_{\bf k}(t)  \tilde{\psi}_{\bf
{k^\prime}}(t^\prime) \rangle \equiv Q^-({\bf k},t,t^\prime)
\delta_{{\bf k} - {\bf k^\prime}},
 \label{eq:18.2}
\end{equation}
and the Green functions $G^\pm({\bf k},t-t^\prime)$. In Eq.
(\ref{eq:18.1}) $\tilde{\phi}_{\bf k}(t)$ is the inverse Fourier
transform of $\phi_{\bf \bar k}$ with respect to $\omega$. For
stationary turbulence $Q^\pm({\bf k},t,t^\prime)=Q^\pm({\bf
k},t-t^\prime)$. Close analogy between the WCA and the DIA could be
seen if one notes that according to Eqs. (\ref{eq:7}) and
(\ref{eq:18.1}), $I_{\bf \bar k}^\pm$ are the Fourier transforms of
$Q^\pm({\bf k},t-t^\prime)$ with respect to $t-t^\prime$, whereas
$\Gamma_{\bf \bar k}^\pm/2\pi$ are the Fourier transforms of
$G^\pm({\bf k},t-t^\prime)$ \cite{K65}.

\subsection{The conservation laws}

It can be checked that Eqs. (\ref{eq:17.1})-(\ref{eq:17.2})
conserves total energy of the both types of the shear Alfv\'en
waves. Indeed, performing the inverse Fourier transform of Eqs.
(\ref{eq:17.20})-(\ref{eq:17.2}) with respect to $\omega$, setting
the temporal variable equal to zero, integrating both sides of the
obtained equations over the whole ${\bf k}$ space and taking into
account that $I_{\bf \bar k}^\pm \equiv I_{-\bf \bar k}^\pm$ it can
be shown that considered nonlinear interactions conserve total
energy of the both types of the shear Alfv\'en waves $H^+=1/2 \int
|\phi|^2{\rm d}^4 {\bar k}$ and $H^-=1/2 \int |\psi|^2{\rm d}^4
{\bar k}$, or equivalently, Eqs. (\ref{eq:17.20})-(\ref{eq:17.2})
conserve both the total energy $H=H^+ + H^-$ and the cross helicity
$H=H^+ - H^-$.

\section{The WCA for isotropic hydrodynamic turbulence} \label{sec:5}

Before performing analysis of the WCA equations
(\ref{eq:17.1})-(\ref{eq:17.2}) for incompressible MHD turbulence,
in this section we study much more simple problem - the WCA
equations for isotropic hydrodynamic turbulence. There are two
reasons to perform this study. Firstly, although WCA equations for
isotropic hydrodynamic turbulence have been derived by different
authors \cite{K65,MY}, no methods of the analysis have been
indicated. The study of these equations allows us to develop the
method for analysis of the WCA equations for the simplest example -
the inertial range of isotropic hydrodynamic turbulence. Secondly,
we derive Kraichnan's nonlocal spectrum. Comparison of the results
obtained in this section with the results for inertial range of
incompressible MHD turbulence obtained in the next sections allow us
to show that the energy spectrum $E_q \sim q^{-3/2}$ observed in the
numerical simulations of incompressible MHD turbulence in a strong
external magnetic field represents the anisotropic analogue of
Kraichnan's nonlocal spectrum.

The equations analogous to (\ref{eq:15.1}) and (\ref{eq:17.20}) for
the inertial range of the hydrodynamic turbulence have the form
\cite{K65,MY}
\begin{equation}
i\zeta_{\bf \bar k}^h = -\int_{-\infty}^\infty \frac{k^2 b_{1,2}
}{\omega_1 + i\nu k_1^2 + i\zeta_1^h} I_2^h d{\cal F}_{1,2}^k.
\label{eq:19.1}
\end{equation}
\begin{eqnarray}
-i\left( \omega + i\nu k^2 \right) I_{\bf \bar k} = k^2 \Gamma_{\bf
\bar k}^{h\ast} \int_{-\infty}^\infty a_{1,2} I_1^h I_2^h d{\cal
F}_{1,2}^k - \nonumber
\\ k^2 I_{\bf \bar k}^h \int_{-\infty}^\infty b_{1,2} \Gamma_1^h I_2^h
d{\cal F}_{1,2}^k,
 \label{eq:19.2}
\end{eqnarray}
where
\begin{eqnarray}
a_{1,2} = \frac{1}{2} \left[ 1-2\frac{({\bf k} \cdot {\bf k}_1 )^2
({\bf k} \cdot {\bf k}_2 )^2 }{k^2 k_1^2 k_2^2} + \right. \nonumber
\\ \left. \frac{({\bf k}
\cdot {\bf k}_1)({\bf k} \cdot {\bf k}_2)({\bf k}_1 \cdot {\bf
k}_2)}{k^2 k_1^2 k_2^2} \right], \label{eq:20.1}
\end{eqnarray}
\begin{equation}
b_{1,2} =  \frac{({\bf k} \cdot {\bf k}_1 )^3 }{k^4 k_1^2} -
\frac{({\bf k} \cdot {\bf k}_2)({\bf k}_1 \cdot {\bf k}_2)}{k^2
k_2^2}, \label{eq:20.2}
\end{equation}
$\zeta_{\bf \bar k}^h$ and $\Gamma_{\bf \bar k}^h$ are related by
the relation (\ref{eq:16}) with $\omega_{\bf k}=-i\nu k^2$, and
$I_{\bf \bar k}^h$ is defined by $\langle {\bf v}_{\bf \bar k} \cdot
{\bf v}_{\bf \bar k^\prime} \rangle =I_{\bf \bar k}^h \delta_{{\bf
\bar k} - {\bf \bar k^\prime}} $, where ${\bf v}_{\bf \bar k}$ is
the Fourier transform of the turbulent velocity field.

Eqs. (\ref{eq:19.1})-(\ref{eq:20.2}) are useless until some
assumptions are made about the frequency dependence of $\Gamma_{\bf
\bar k}^h$ and $I_{\bf \bar k}^h$. Equivalently, in the framework of
the DIA one should make some assumptions about the time dependence
of $G^h({\bf k},t-t^\prime)$ and $Q^h({\bf k},t-t^\prime)$
\cite{Le}, which are the corresponding inverse Fourier transforms
with respect to $\omega$. One of the simplest and frequently used
assumptions imply \cite{K59,Le}
\begin{equation}
G^h({\bf k},t-t^\prime)=\exp{[-|\eta_{\bf k}^h|(t-t^\prime)]}
H(t-t^\prime), \label{eq:21.1}
\end{equation}
\begin{equation}
Q^h({\bf k},t-t^\prime)=\exp{[-|\xi_{\bf k}^h|(t-t^\prime)]}
\mathcal{E}_{\bf k}, \label{eq:21.2}
\end{equation}
where $H(t)$ is the Heaviside (step) function, and $\mathcal{E}_{\bf
k}$ is the energy spectrum. Usually it is also assumed \cite{E64,Le}
that the temporal autocorrelation scales of the spectral function
and the Green function are equal $1/\eta_{\bf k}^h=1/\xi_{\bf k}^h
\equiv \tau_{ac}^h$. From Eq. (\ref{eq:21.2}) it follows that
temporal derivative of $Q^h({\bf k},t-t^\prime)$ has the
discontinuity at $\tau \equiv t-t^\prime=0$. As it was shown in Ref.
\cite{E64} this makes no problems in analysis of the turbulence
dynamics if one assumes
\begin{equation}
\lim_{\tau \rightarrow 0} \frac{\rm d}{{\rm d} \tau} Q^h({\bf
k},\tau)=0. \label{eq:21.3}
\end{equation}

Similar to Eqs. (\ref{eq:21.1})-(\ref{eq:21.2}), in the case under
consideration we assume
\begin{equation}
\zeta_{\bf \bar k}^h=\eta_{\bf k}^h, \label{eq:22.1}
\end{equation}
\begin{equation}
I_{\bf \bar k}^h = \frac{\mathcal{E}_{\bf k}}{\pi} \frac{\eta_{\bf
k}^h}{\omega^2 + ({\eta_{\bf k}^h})^2}, \label{eq:22.2}
\end{equation}
Substituting these equations into Eq. (\ref{eq:19.1}) and
integrating over the frequency variables we obtain
\begin{equation}
\eta_{\bf \bar k}^h = \int_{-\infty}^\infty  \frac{i k^2 b_{1,2}
\mathcal{E}_{{\bf k}_2}}{\omega+i(\eta_{{\bf k}_1}^h + \eta_{{\bf
k}_2}^h)} {\rm d} \mathcal{K}_{1,2}^k, \label{eq:23}
\end{equation}
where ${\rm d} \mathcal{K}_{1,2}^k \equiv {\rm d}^3{\bf k}_1 {\rm
d}^3{\bf k}_2 \delta_{{\bf k}- {\bf k}_1 -{\bf k}_2}$. As we see the
right hand side of Eq. (\ref{eq:23}) depends on $\omega$, whereas
according to our assumption the left hand side does not. This means,
that Eqs. (\ref{eq:22.1}) and (\ref{eq:22.2}) can not be valid for
the whole $({\bf k},\omega)$ space. The solution of this
inconsistency is based on the following physical arguments:
introducing the autocorrelation time scale $\tau_{ac}=1/\eta_{\bf
k}^h$ one also should assume that $\tau_{ac}$ is the shortest time
scale in the problem and the dynamics of turbulence is totally
determined by $\tau_{ac}$. Alternatively, it can be shown that the
main contribution in the energy balance equation (\ref{eq:19.2})
comes from the modes with $\omega \lesssim \eta_{\bf k}$, and
therefore when studying the scaling properties of the inertial range
of the turbulence one can set $\omega=0$ in Eq. (\ref{eq:23}).
Taking also into account the isotropy of the turbulence ($\eta_{\bf
k}^h=\eta_k^h$ and $\mathcal{E}_{\bf k}^h=\mathcal{E}_k^h$) we
obtain
\begin{equation}
\eta_k^h = k^2 \int_{-\infty}^\infty b_{1,2} \frac{
\mathcal{E}_{k_2}}{\eta_{k_1}^h + \eta_{k_2}^h} {\rm d}
\mathcal{K}_{1,2}^k. \label{eq:24}
\end{equation}

It should be noted, that the standard DIA technique (see, e.g., Ref.
\cite{Le}) also leads to Eq. (\ref{eq:24}) without explicit
assumption that $\tau_{ac}$ is the only timescale which totally
determines the dynamics of turbulence in the inertial interval. As
it is shown in the Appendix, the method of derivation of Eq.
(\ref{eq:24}) in the framework of the DIA also implies implicitly
that the formulated assumption is hold.

For analysis of Eq. (\ref{eq:19.2}) we perform inverse Fourier
transform with respect to $\omega$. In the right hand side of the
obtained equation we can set $\nu=0$, due to the fact that the
dissipation range has negligible influence on the nonlinear energy
transfer in the inertial range. Substituting Eqs.
(\ref{eq:22.1})-(\ref{eq:22.2}), performing integration with respect
to frequencies, using Eq. (\ref{eq:21.3}) and taking into account
the identity \cite{Le}
\begin{equation}
a_{1,2}=\frac{1}{2}\left( b_{1,2}+ b_{2,1} \right). \label{eq:25}
\end{equation}
we finally obtain
\begin{equation}
\nu k^2 \mathcal{E}_k = k^2 \int_{-\infty}^\infty b_{1,2} \frac{
\mathcal{E}_{k_2}(\mathcal{E}_{k_1}-\mathcal{E}_{k})}{\eta_{k}^h+\eta_{k_1}^h
+ \eta_{k_2}^h} {\rm d} \mathcal{K}_{1,2}^k. \label{eq:26.0}
\end{equation}
For further simplification of this equation we multiply it by $4\pi
k^2$ and integrate from some $k$ in the inertial interval to the
infinity. For isotropic hydrodynamic turbulence the effective
dissipation takes place for high wave numbers and therefore
\begin{equation}
4\pi \nu \int_{k}^\infty k^4 \mathcal{E}_k {\rm d}k \approx \nu
\int_{-\infty}^\infty k^2 \mathcal{E}_k {\rm d}^3 {\bf k}
=\varepsilon_h, \label{eq:27.0}
\end{equation}
where $\varepsilon_h$ is the energy dissipation rate. Then Eq.
(\ref{eq:26.0}) reduces to
\begin{equation}
\varepsilon_h = 4\pi \int_k^\infty k^4 {\rm d} k
\int_{-\infty}^\infty \frac{b_{1,2}
\mathcal{E}_{k_2}(\mathcal{E}_{k_1}-\mathcal{E}_{k})}{\eta_{k}^h+\eta_{k_1}^h
+ \eta_{k_2}^h} {\rm d} \mathcal{K}_{1,2}^k. \label{eq:26}
\end{equation}

To study the scaling properties of the turbulence we seek for the
inertial range solution of Eqs. (\ref{eq:24}) and (\ref{eq:26.0}) in
the form
\begin{equation}
\mathcal{E}_{k}=A_h k^m, \label{eq:27}
\end{equation}
\begin{equation}
\eta_k^h=B_h k^n. \label{eq:28}
\end{equation}

Substituting these expressions into Eq. (\ref{eq:24}) and taking
into account that for $k \gg k_2$, $b_{1,2}\approx \sin^2\alpha_2$,
where $\alpha_2$ is the angle between ${\bf k}$ and ${\bf k}_2$, we
see that the integral on the right hand side of Eq. (\ref{eq:24})
diverges at $k_2 \rightarrow 0$ if $m< -3$ and $n>0$. Note that for
the Kolmogorov spectrum $m=-11/3$ and $n=2/3$. Therefore, the
integral in Eq. (\ref{eq:24}) is dominated by the low wave number
energy containing modes for which Eq. (\ref{eq:27}) is not valid.

To calculate the contribution of the energy containing modes we
substitute Eq. (\ref{eq:27}) into Eq. (\ref{eq:24}) and use delta
function to integrate over $k_1$. Noting that $|{\bf k}-{\bf
k}_2|\approx k$ this yields
\begin{equation}
B_h k^n \approx B_h^{-1} k^{2-n} \int \mathcal{E}_{k_2} \sin^2
\alpha_2 {\rm d}^3 {\bf k}_2 \sim B_h^{-1} k^{2-n} v_0^2,
\label{eq:29}
\end{equation}
where $v_0$ is the characteristic velocity of energy containing
vortices. From Eq. (\ref{eq:29}) we obtain
\begin{equation}
n=1, ~~~~~~~~~~B_h \sim v_0 \label{eq:30}
\end{equation}

In contrast to Eq. (\ref{eq:24}), due to the presence of the
multiplier $(\mathcal{E}_{k_1}-\mathcal{E}_k)$, integrals in Eq.
(\ref{eq:26}) are convergent for the small values of $k_2$ when
$m>-4$. Consequently, noting that the main contribution in the
integral comes from the wave numbers with $k_1\sim k_2 \sim k$ we
obtain
\begin{equation}
\varepsilon_h \sim A_h^2 B_h^{-1} k^{8+2m-n}, \label{eq:31}
\end{equation}
and therefore $8+2m-n=0$ and $A_h \sim (\varepsilon_h B_h)^{1/2}$.
Using Eqs. (\ref{eq:30}) and introducing the one dimensional energy
spectrum $E_k =4\pi k^2 \mathcal{E}_k$ we obtain
\begin{equation}
E_k \sim (\varepsilon_h v_0)^{1/2} k^{-3/2}, ~~~ \eta^h_k \sim k
v_0, \label{eq:32}
\end{equation}
which represents Kraichnan's nonlocal spectrum \cite{K59}. Note that
in the considered model the spectral energy transfer is local [i.e.,
integrals in Eq. (\ref{eq:26}) are convergent] and the nonlocal
character of the turbulence is related to effective decorrelation
caused by low frequency modes from energy containing interval
[divergence of the integrals in Eq. (\ref{eq:24}) and as a
consequence shortening of the autocorrelation time scale from
$1/(kv_l)$ to $1/(kv_0)$].

Solution of the contradiction between Eq. (\ref{eq:32}) and
Kolmogorov spectrum led Kraichnan to the formulation of so-called
Lagrangian DIA \cite{Kr65}. Another approach was developed by
Kadomtsev \cite{K65}. He argued that analysis of Eq. (\ref{eq:24})
which led to Eq. (\ref{eq:29}) and (\ref{eq:30}) overestimates the
contribution of nonlocal interactions. It was asserted that when
studying nonlinear interactions of the modes, one should distinguish
two kinds of nonlinear interactions - resonant and adiabatic. In Eq.
(\ref{eq:24}) it is implicitly assumed that all the nonlinear
interactions are resonant. In fact, interactions of the modes with
very different wave numbers are adiabatic rather then resonant: the
influence of the large scale fluctuations on the low frequency ones
leads to adiabatic (WKB) change of the frequency $\omega$ and the
wave number ${\bf k}$ of the high frequency fluctuation, similar to
the weak inhomogeneity of the mean fields.

As it was shown in Ref. \cite{K65}, the dynamics of the turbulence
in the inertial range is mainly determined by the resonant nonlinear
interactions whereas influence of the adiabatic interactions can be
neglected. From mathematical point of view this means, that instead
of integration over the whole ${\bf k}$-space in Eq. (\ref{eq:24}),
one should restrict the area of integration by the area $ \sigma k <
|{\bf k}_{1,2}| <k/\sigma$, where $\sigma \sim 1$, say $\sigma=1/3$,
\cite{K65} or $\sigma=1/\sqrt{2}$, \cite{Le,B03}. If so, using Eqs.
(\ref{eq:27})-(\ref{eq:28}), instead of Eq. (\ref{eq:29}) we obtain
\begin{equation}
Bk^n \sim A B^{-1} k^{5+m-n}. \label{eq:35}
\end{equation}
Combining this equation with Eq. (\ref{eq:31}) we arrive at the
Kolmogorov spectrum $E_{k} \sim \varepsilon_h^{2/3} k^{-5/3}$. Note
that for the Kolmogorov spectrum $\tau_{ac}^K \sim \tau_{cas}^K \sim
1/kv_l$.

As it will be shown in the Sec. \ref{sec:7}, although the concept of
resonant and adiabatic interactions yields right result for inertial
range of isotropic hydrodynamic turbulence, for the systems with
more then one degrees of freedom it does not hold.

\section{Nonlocal spectrum of the incompressible MHD turbulence} \label{sec:6}

For further analysis of Eqs. (\ref{eq:15.1})-(\ref{eq:15.2}), or
equivalently  (\ref{eq:17.1})-(\ref{eq:17.2}), similar to the
hydrodynamic case we assume
\begin{equation}
G^\pm({\bf k},\tau)=\exp{\left( -|\eta_{\bf k}^\pm| \tau \pm i
\omega_{\bf k} \tau \right)} H(\tau), \label{eq:36.1}
\end{equation}
\begin{equation}
Q^\pm({\bf k},\tau)=\exp{\left(-|\xi_{\bf k}^\pm|\tau \pm i
\omega_{\bf k} \tau \right)} \mathcal{E}_{\bf k}, \label{eq:36.2}
\end{equation}
or equivalently
\begin{equation}
\zeta_{\bf \bar k}^\pm=\eta_{\bf k}^\pm, \label{eq:37.1}
\end{equation}
\begin{equation}
I_{\bf \bar k}^\pm = \frac{\mathcal{E}_{\bf k}^\pm}{\pi}
\frac{\xi_{\bf k}^\pm}{(\omega \mp \omega_{\bf k})^2 + ({\xi_{\bf
k}^\pm})^2}. \label{eq:37.2}
\end{equation}
In the presented paper we consider the symmetric case
\begin{equation}
\eta_{\bf k}^\pm = \xi_{\bf k}^\pm \equiv \eta_{\bf
k},~~~\mathcal{E}_{\bf k}^+ = \mathcal{E}_{\bf k}^- \equiv
\mathcal{E}_{\bf k}, \label{eq:38}
\end{equation}
which corresponds to the turbulence with zero cross helicity
($\mathcal{E}_{\bf k}^+ - \mathcal{E}_{\bf k}^- =0$). In this case
$I_{{\bf k},\omega}^\pm = I_{{\bf k},-\omega}^\mp$, and it can be
readily checked that Eqs. (\ref{eq:15.1}) and (\ref{eq:15.20})
coincides with Eqs. (\ref{eq:15.10}) and (\ref{eq:15.2})
respectively. Therefore, in the symmetric case there remain only two
independent equations.

Here we consider strong turbulence, i.e., assume that the
autocorrelation timescale $\tau_{ac}=1/\eta_{\bf k}$ is the shortest
timescale presented in the problem which totally determines the
dynamics of the turbulence. Taking this into account and performing
integrations with respect to frequencies in Eqs. (\ref{eq:15.1}) and
(\ref{eq:17.20}) similar to hydrodynamic turbulence we obtain
\begin{equation}
\eta_{\bf k} = \int_{-\infty}^\infty |T_{1,2}|^2 \frac{
\mathcal{E}_{{\bf k}_2}}{\eta_{{\bf k}_1} + \eta_{{\bf k}_2}} {\rm
d} \mathcal{K}_{1,2}^k, \label{eq:39}
\end{equation}
\begin{equation}
\bar \nu k^2 \mathcal{E}_{\bf k} = \int_{-\infty}^\infty |T_{1,2}|^2
\frac{ \mathcal{E}_{{\bf k}_2} (\mathcal{E}_{{\bf
k}_1}-\mathcal{E}_{{\bf k}})}{\eta_{\bf k}+\eta_{{\bf k}_1} +
\eta_{{\bf k}_2}} {\rm d} \mathcal{K}_{1,2}^k. \label{eq:40}
\end{equation}
For analysis of anisotropic turbulence we assume that the dynamics
is dominated by cascade perpendicular with respect to the mean
magnetic field and therefore we assume $\eta_{\bf k} = \eta_{q}$.
For energy spectrum we assume \cite{GS95}
\begin{equation}
\mathcal{E}_{\bf k} = \frac{A}{\Lambda} q^{m-\mu} f\left(
\frac{p}{\Lambda q^\mu} \right), \label{eq:41}
\end{equation}
where $f(u)$ is a positive symmetric function of $u$ which becomes
negligibly small for $u \gg 1$. For $u \lesssim 1$, $f(u) \sim 1$
such that $\int_{-\infty}^{\infty} f(u)du=1$. We also define the two
dimensional energy density as
\begin{equation}
\bar \mathcal{E}_{\bf q} \equiv \int_{-\infty}^{\infty}
\mathcal{E}_{\bf k} {\rm d}p = Aq^m. \label{eq:42}
\end{equation}
Performing the integration with respect to $p_1$ and $p_2$ in Eqs.
(\ref{eq:39})-(\ref{eq:40}), multiplying both sides of Eq.
(\ref{eq:40}) by $2\pi q$ and integrating over $p$ and $q$ from some
value in the inertial range to infinity and defining the energy
dissipation rate
\begin{equation}
\varepsilon \equiv {\bar \nu} \int_{-\infty}^\infty k^2
\mathcal{E}_{\bf k} {\rm d}^3 {\bf k} = 2\pi {\bar \nu}
\int_{0}^{\infty} q^3 \bar \mathcal{E}_{\bf q} {\rm d}q.
\label{eq:43}
\end{equation}
we obtain
\begin{equation}
\eta_{q} = \int_{-\infty}^\infty |T_{1,2}|^2 \frac{ \bar
\mathcal{E}_{{\bf q}_2}}{\eta_{q_1} + \eta_{q_2}} {\rm d}
\mathcal{Q}_{1,2}^k, \label{eq:44}
\end{equation}
\begin{equation}
\varepsilon = 2\pi \int_q^\infty q{\rm d} q \int_{-\infty}^\infty
|T_{1,2}|^2 \frac{ \bar \mathcal{E}_{{\bf q}_2} (\bar
\mathcal{E}_{{\bf q}_1}-\bar \mathcal{E}_{{\bf
q}})}{\eta_q+\eta_{q_1} + \eta_{q_2}} {\rm d} \mathcal{Q}_{1,2}^k,
\label{eq:45}
\end{equation}
where ${\rm d} \mathcal{Q}_{1,2}^k \equiv {\rm d}^2{\bf q}_1 {\rm
d}^2{\bf q}_2 \delta_{{\bf q}- {\bf q}_1 -{\bf q}_2}$.

Similar to Eq. (\ref{eq:28}) we assume $\eta_q =Bq^n$. Taking into
account that $|T_{1,2}|^2 \equiv q^2 \cos^2{\theta_1}
\sin^2\theta_2$, where $\theta_{1,2}$ are the angles between $\bf q$
and ${\bf q}_{1,2}$ respectively, we obtain that the integral on the
right hand side of Eq. (\ref{eq:44}) is divergent for $q_2
\rightarrow 0$ if $m<-2$. Further analysis is analogous to one
performed in Sec. \ref{sec:5} for hydrodynamic turbulence. Similarly
we obtain:
\begin{equation}
{\bar \mathcal{E}}_{\bf q} \sim (\varepsilon v_0)^{1/2}
q^{-5/2},~~~~\eta_q \sim qv_0. \label{eq:46}
\end{equation}
For the one dimensional spectrum $E_q \equiv 2\pi q {\bar
\mathcal{E}}_{\bf q}$ this yields $E_q \sim (\varepsilon v_0)^{1/2}
q^{-3/2}$. For the characteristic cascade timescale we obtain:
\begin{equation}
\tau_{cas} \sim \frac{1}{qv_q} \frac{v_0}{v_q}. \label{eq:47}
\end{equation}

The analysis of the anisotropic turbulence is not full until some
relation between parallel and perpendicular length scales of the
turbulent wave packets are specified. Technically this implies
determination of $\mu$ and $\Lambda$. This could be done based on
arguments discussed in Sec. \ref{sec:2}, and lead us to Eq.
(\ref{eq:0.10}) or equivalently
\begin{equation}
\mu=1/2,~~~~~ \Lambda = \frac{v_0}{V_A} q_0^{1/2}. \label{eq:50}
\end{equation}

\section{Random Galilean invariance and locality of incompressible MHD turbulence} \label{sec:7}

The study of the locality of incompressible MHD turbulence performed
in this section is based on the methods developed in Ref.
\cite{K65}. Let us isolate resonant interactions in Eq.
(\ref{eq:6.1}) and represent this equation in the following form
\begin{eqnarray}
(\omega-\omega_{\bf k}) \phi_{\bf \bar k}= \int_{\mathcal{R}}
T_{1,2} \phi_{1} \psi_{2} {\rm d} {\cal F}_{1,2}^k +
\int_{\mathcal{D}} T_{1,2} \phi_{1} \psi_{2} {\rm d} {\cal
F}_{1,2}^k + \nonumber
\\ \int_{\mathcal{A}_1} T_{1,{\bf k}-1} \phi_{1} \psi_{{\bf k}-1} {\rm d}^3{\bf k}_1 +
\int_{\mathcal{A}_2} T_{{\bf k}-2,2} \phi_{{\bf k}-2} \psi_{2} {\rm
d}^3{\bf k}_2.
 \label{eq:51}
\end{eqnarray}
In this equation $\mathcal R$ denotes the resonant area where $q_1
\sim q_2 \sim q$ (as in the previous section we also assume that
there exist some relation between the perpendicular and parallel
length scales of turbulent wave packets \cite{GS95}), $\mathcal D$
denotes the high frequency area where $q_1 \sim q_2 \gg q$, and
${\mathcal A}_{1,2}$ denote the areas $q_{1,2} \ll q$, respectively.
The analysis performed in the previous section shows that the area
$\mathcal D$ has negligible contribution to the integrals in Eqs.
(\ref{eq:39})-(\ref{eq:40}) and will be neglected in the further
consideration. However, it should be noted, that nonlocal
interactions with small scale fluctuations could have important
contributions in the dynamics of MHD turbulence with nonzero
helicity \cite{B03,MG05}.

Performing Taylor expansion in the integrands of the third and forth
terms on the right hand side of Eq. (\ref{eq:51}) with respect to
the small parameter $q_{1,2}/q$ respectively and taking into account
the definition of $T_{1,2}$ and Eq. (\ref{eq:5}) we obtain
\begin{eqnarray}
(\omega-\omega_{\bf k}) \phi_{\bf \bar k}= {\bf k}\cdot({\bf V}_L +
{\bf B}_L) \phi_{\bf \bar k} + G_{\phi \phi} + G_{\phi \psi}+
\nonumber \\ \int_{\mathcal{R}} T_{1,2} \phi_{1} \psi_{2} {\rm d}
{\cal F}_{1,2}^k.
 \label{eq:52}
\end{eqnarray}
Similar manipulations of Eq. (\ref{eq:6.2}) yields
\begin{eqnarray}
(\omega+\omega_{\bf k}) \psi_{\bf \bar k}= {\bf k}\cdot({\bf V}_L -
{\bf B}_L) \psi_{\bf \bar k} + G_{\psi \phi} + G_{\psi \psi} +
\nonumber \\ \int_{\mathcal{R}} T_{1,2} \psi_{1} \phi_{2} {\rm d}
{\cal F}_{1,2}^k.
 \label{eq:53}
\end{eqnarray}
In Eqs. (\ref{eq:52}) and (\ref{eq:53}) $G_{ij}$ are the First order
terms of the Taylor expansion $G_{i\phi}\sim \partial \phi /\partial
{\bf k}$ and $G_{i\psi}\sim \partial \psi /\partial {\bf k}$. ${\bf
V}_L=\int_{k_1\ll k} {\bf v}_{{\bf \bar k}_1} d^4{\bf \bar k}_1$ and
${\bf B}_L\int_{k_1\ll k} {\bf b}_{{\bf \bar k}_1} d^4{\bf \bar
k}_1$ are the velocity and the magnetic fields of the low frequency
modes.

According to the principle of resonant and adiabatic interactions,
first terms of the right hand sides of Eqs. (\ref{eq:52}) and
(\ref{eq:53}) describe the transfer of high frequency perturbations
by the low frequency modes and one should eliminate these terms
before applying the closure procedure described in Sec. \ref{sec:4}.
In the case under consideration this could be done as follows: the
terms proportional to ${\bf V}_L$ can be eliminated by means of the
Galilean transformation (this will lead to the frequency
renormalization $\omega^\prime=\omega-{\bf k}\cdot {\bf V}_L$),
whereas the terms proportional to ${\bf B}_L$ can be included to
$\omega_{\bf k}$. If so, the influence of low frequency modes on the
dynamics of the high frequency modes comes to slow (adiabatic)
deformation of the high frequency modes, described by $G_{ij}$ terms
in Eqs. (\ref{eq:52}) and (\ref{eq:53}). In the zeroth order
approximation the adiabatic interactions can be neglected and
therefore Eqs. (\ref{eq:52})-(\ref{eq:53}) formally coincide with
Eqs. (\ref{eq:6.1})-(\ref{eq:6.2}), except the integration now is
performed only over the resonant area $\mathcal R$. Applying the WCA
(DIA) closure scheme described in Sec. \ref{sec:4} to these
equations, we obtain Eqs. (\ref{eq:17.1})-(\ref{eq:17.2}) with the
integration performed only over the resonant area $\mathcal R$. Then
analysis similar to the performed in the previous section would lead
to the GS model of the turbulence described by Eqs. (\ref{eq:0.4}).

The WCA (Eulerian DIA) suggests that low frequency fluctuations
always cause decorrelation of interactions of high frequency
fluctuations and as it is known this result is wrong
\cite{K65,Kr65}. In contrary, the principle of resonant and
adiabatic interactions states that the influence of low frequency
modes on the dynamics of high frequency modes always have adiabatic
character. Although this principle yields right results for inertial
range of isotropic hydrodynamic turbulence as well as for any system
with one degree of freedom [i.e., when there exist one starting
equation similar to Eqs. (\ref{eq:6.1})-(\ref{eq:6.2})], in general
case this principle is not correct.

Indeed, in the case of the velocity field the necessity of
elimination of the terms proportional to ${\bf V}_L$ from Eqs.
(\ref{eq:52})-(\ref{eq:53}) is caused by the requirement of the
invariance of the governing equations with respect to the random
Galilean transformation \cite{Kr65}. According to this principle the
'right' form of Eqs. (\ref{eq:52})-(\ref{eq:53}) has ${\bf V}_L=0$.
In the case of magnetic field there exists no such a fundamental
principle and consequently the necessity of the elimination of the
term proportional to ${\bf B}_L$ is not obvious. Physical picture is
the same as it was discussed in Sec. \ref{sec:3}: If the low
frequency energy containing modes contribute to the mean magnetic
field acting on the high frequency modes, then the terms
proportional to ${\bf B}_L$ should be included in terms proportional
to $\omega_{\bf k}$ on the left hand side of Eqs.
(\ref{eq:52})-(\ref{eq:53}). In this case the DIA closure scheme
presented in Sec. \ref{sec:4} and analysis similar to one presented
in Sec. \ref{sec:6} leads to the GS model of the incompressible MHD
turbulence. On the other hand, if the low frequency modes do not
contribute to the mean magnetic field, then the same kind of
analysis leads to the anisotropic analogue of Kraichnan's nonlocal
spectrum. Therefore, we conclude that in contrary to the
incompressible hydrodynamic turbulence, incompressible MHD
turbulence could be both local and nonlocal.

In the general case it is clear that the principle of resonant and
adiabatic interactions holds for any system with one degree of
freedom, since in this case the zeroth order term of the Taylor
expansion describing influence of the low frequency modes on the
high frequency modes always can be eliminated by corresponding
Galilean transformation. In the case of the system with two or more
degrees of freedom random Galilean invariance can not guarantee the
elimination of all zeroth order terms, as in the considered above
case of Eqs. (\ref{eq:52})-(\ref{eq:53}) and therefore the low
frequency modes can cause effective decorrelation of high frequency
fluctuations.

\section{Two dimensional MHD turbulence} \label{sec:71}

It is well known that dynamics of two dimensional hydrodynamic
turbulence is basically different from three dimensional turbulence.
The origin of this difference is  that in two dimensional case in
addition to the energy there exist another conserved quantity -
enstrophy. The situation is different in magnetohydrodynamics. As it
was discussed above, MHD turbulence is highly anisotropic, dominated
by turbulent cascade in the direction perpendicular to the mean
magnetic field. Therefore it has been argued that the local two
dimensional dynamics in the plane perpendicular to the mean magnetic
field describe an essential part of the three dimensional turbulence
(especially in the presence of the strong mean field) and could be
studied by two dimensional MHD. In this context two dimensional MHD
turbulence has been extensively studies by different authors
\cite{BW89,PPC98,BS01}. It seems interesting to compare presented
model with the features of two dimensional MHD turbulence observed
in numerical simulations. Numerical simulations performed in Refs.
\cite{BW89,BS01} show that inertial range spectrum of two
dimensional MHD turbulence in the absence of a mean magnetic field
has the form
\begin{equation}
E_k = C_k \varepsilon^{1/2} E_M^{1/4} k^{-3/2}, \label{eq:511}
\end{equation}
with $C_k\approx 1.8$ and $E_M$ is the density of the magnetic
energy. Noting that $E_M \sim v_0^2$ we conclude that this result
coincides with the prediction of the presented model for
perpendicular cascade given by Eq. (\ref{eq:0.8}). Although this
coincidence is quite remarkable, it can not be treated as
confirmation of the presented model, because it is not obvious that
two dimensional MHD properly reproduces local structure of the three
dimensional MHD turbulence in the place perpendicular to the mean
magnetic field. It also should be noted that the results obtained in
Ref. \cite{PPC98} for the inertial range of two dimensional MHD
turbulence are closer to the Kolmogorov spectrum.

On the other hand, two dimensional MHD represents the system with
two degrees of freedom (see, e.g., Eqs. (1)-(2) of Ref. \cite{BW89})
which keep several important properties of MHD. Namely, linear modes
have the same dispersion and Alfv\'en effect is also presented in
two dimensional MHD. In this context theoretical as well as
numerical study (including the case with strong background magnetic
field) of two dimensional MHD seems to have its own value,
especially for the study of the turbulence locality.

\section{Conclusions} \label{sec:8}

Novel model of incompressible magnetohydrodynamic turbulence in the
presence of a strong external magnetic field is proposed. It is
suggested that in the presence of the strong external magnetic field
incompressible MHD turbulence becomes nonlocal in the sense that the
low frequency modes cause decorrelation of interacting high
frequency modes from the inertial interval. Obtained nonlocal
spectrum of the inertial range of incompressible MHD turbulence,
given by Eqs. (\ref{eq:0.8}) and (\ref{eq:0.10}), represents the
anisotropic analogue of Kraichnan's nonlocal spectrum of isotropic
hydrodynamic turbulence. Based on the analysis performed in the
framework of WCA it is shown that in contrast to the isotropic
hydrodynamic turbulence, incompressible MHD turbulence could be both
local and nonlocal and therefore an anisotropic analogues of both
Kolmogorov and Kraichnan spectra are realizable in incompressible
MHD turbulence.

\begin{acknowledgments}
I am grateful to Mariam Akhalkatsi, Wendell Horton,  George
Machabeli and Wolf-Christian M\"{u}ller for helpful discussions and
suggestions. This research was supported by Georgian NSF grant
ST06/4-096 and INTAS grant 06-1000017-9258.

\end{acknowledgments}

\appendix
\section{}

DIA equation that governs the dynamics of the Green function
$G^h({\bf k},t)$ for the inertial range of the isotropic
hydrodynamic turbulence has the form \cite{K59,Le}
\begin{eqnarray}
\frac{{\rm d}}{{\rm d} t} G^h(k,t-t^\prime) = -
\int_{-\infty}^{\infty} {\rm d}^3 {\bf k}_1 {\rm d}^3 {\bf k}_2
\delta_{{\bf k}-{\bf k}_1-{\bf k}_2} b_{1,2} \times \nonumber \\
\int_{t^\prime}^{t} {\rm d} t^{\prime \prime} G^h(k_1,t-t^{\prime
\prime}) Q^h(k_2,t-t^{\prime \prime}) G^h(k,t^{\prime
\prime}-t^\prime). \label{eq:A1}
\end{eqnarray}

The standard method of the analysis \cite{Le} implies integration of
this equation with respect to $\tau=t-t^\prime$ in the interval
$(0,\infty)$. Taking also into account Eqs.
(\ref{eq:21.1})-(\ref{eq:21.2}), after the straightforward
manipulations this leads to Eq. (\ref{eq:24}). Described method of
derivation of Eq. (\ref{eq:24}) is not unique. Before performing
integration with respect to $\tau$ we can multiply both sides of Eq.
(\ref{eq:A1}) by any function of $t$. For instance, if we multiply
Eq. (\ref{eq:A1}) by $\exp{(- \alpha t)}$ and then repeat the same
procedures, we obtain Eq. (\ref{eq:23}) with $\omega$ replaced by
$i\alpha$. The reason why this equation is 'wrong', whereas Eq.
(\ref{eq:24}) is 'right' is the same as in WCA analysis presented in
Sec. \ref{sec:5}: it is implied that $\tau_{ac}$ is the shortest
timescale presented in the problem which totally determines the
dynamics of the turbulence.



\end{document}